\documentclass[preprint,superscriptaddress,prc,aps,amssymb,amsfonts]{revtex4}
\usepackage{graphicx}
\usepackage{dcolumn}
\usepackage{bm}
\usepackage{amsmath}    
\usepackage{graphicx}   

\newcommand{\be}{\begin{equation}}
\newcommand{\ee}{\end{equation}}
\newcommand{\bea}{\begin{eqnarray}}
\newcommand{\eea}{\end{eqnarray}}

\newcommand{\ve}{\varepsilon}

\begin{document}

\title{Quasiparticle-vibration coupling in relativistic framework:
shell structure of Z=120 isotopes}

\author{Elena Litvinova}
\affiliation{ExtreMe Matter Institute EMMI and Research Division,
GSI Helmholtzzentrum f\"ur Schwerionenforschung, Planckstra\ss e 1,
D-64291 Darmstadt, Germany}
\date{\today}

\begin{abstract}
For the first time, the shell structure of open-shell nuclei is
described in a fully self-consistent extension of the covariant
energy density functional theory. The approach implies
quasiparticle-vibration coupling for superfluid systems. One-body
Dyson equation formulated in the doubled quasiparticle space of
Dirac spinors is solved for nucleonic propagators in tin isotopes
which represent the reference case: the obtained energies of the
single-quasiparticle levels and their spectroscopic amplitudes are
in agreement with data. The model is applied to describe the shell
evolution in a chain of superheavy isotopes $^{292,296,300,304}$120
and finds a rather stable proton spherical shell closure at Z = 120.
An interplay of the pairing correlations and the
quasiparticle-phonon coupling gives rise for a smooth evolution of
the neutron shell gap between N = 172 and N = 184 neutron numbers.
Vibrational corrections to the alpha decay energies reach several
hundred keV and can be either positive and negative, thus also
smearing the shell effects.

\end{abstract}
\pacs{21.10.Pc,21.10.Re,21.60.Jz,27.90.+b}

\maketitle

%
%
Impressive progress of experimental low-energy nuclear physics such
as synthesis of many exotic nuclei \cite{GG.08}, superheavy nuclei
\cite{O.06,O.10,H.10} and discovering new nuclear structure
phenomena \cite{A.08} insistently calls for conceptually new
theoretical methods. High-precision description of nuclear
properties still remains a challenge for contemporary theoretical
physics. Besides describing the experimentally known phenomena,
theory has to make predictions to guide future experimental searches
and to model physical situations which are not yet available in
laboratories, but occur in the astrophysical conditions.

One of the most promising strategies for medium-mass and heavy
nuclei is the construction of a "universal" nuclear energy density
functional \cite{UNEDF} supplemented by various many-body
correlations. The existing and commonly used concepts do not yet
allow a high-precision description of nuclear properties due to
their very limited or not self-consistent treatment of many-body
correlations. However, such a reliable description is urgently
needed for fast progressing disciplines like nuclear astrophysics or
synthesis of superheavy elements. Delicate interplay of different
kinds of correlations is responsible for binding loosely-bound
systems, decay properties and for low-energy spectra.

Our recent attempts to extend the covariant energy density
functional (CEDF) approach use the relativistic framework
\cite{R.96,V.05} in combination with advancements of the Landau -
Migdal theory for Fermi liquids in parameter-free quantum field
theory techniques \cite{LR.06,LRT.08,LRT.10}. Couplings of
single-particle and collective degrees of freedom are included far
beyond the standard mean field and random phase approximations in a
fully self-consistent way. On top of the CEDF these techniques
turned out to be very successful in the description of nuclear
low-energy dynamics even for exotic very neutron-rich nuclei
\cite{LRT.08,LRTL.09,LRT.10,E.10,LA.11}. The considerable success of
these self-consistent many-body methods shows that they (i)
represent the right strategy towards a universal and precise
approach and (ii) already allow exploration of experimentally
unknown regions of the nuclear chart: nuclei with exotic neutron to
proton (N/Z) ratios and superheavy nuclei.


In the approaches based on the density functional concept,
single-particle properties such as energies and spectroscopic
amplitudes are the key ingredients for a description of nuclear
masses, decay properties and responses to diverse external fields.
In turn, the latter quantities are an essential part of the nuclear
physics input for astrophysical applications like r-process
nucleosynthesis studies \cite{GLM.07} which require the information
about many nuclei including exotic ones. It has been found recently
that shell structure in nuclei with extreme N/Z ratios deviates from
the usual picture and magic numbers are shown to change as functions
of N and Z \cite{OSH.10}.

The shell structure of superheavy nuclei is another challenge for
microscopic models: to define the location of spherical shell gaps
in this area of the nuclear chart is necessary to determine the
regions of stability of these nuclei. Up to now, there is no
consensus about the spherical shell closures above the proton Z = 82
and neutron N = 126 ones: predictions such as Z = 114, Z = 120 or Z
= 126 for the proton and N = 172 or N = 184 for the neutron magic
numbers can be found in the literature \cite{BRRMG.99,SP.07}. The Z
= 120 and N = 172 shell closures predicted by the relativistic and
some Skyrme mean-field models are found to be related to a central
depression of the nuclear density distribution
\cite{BRRMG.99,AF.05}. The Z = 120 element represents a challenge
for future experimental synthesis since it is located at the limits
of accessibility with available cold fusion reactions. Therefore,
accurate estimations of its characteristics are needed from the
theoretical side.

The predictions made by the mean field models, however, ignore
correlations which can play a significant role in the superheavy
mass region where the expected spherical shell gaps are considerably
smaller (2-3 MeV) than in lighter nuclei and pairing correlations of
the superfluid type may not collapse at the shell closures. It has
been found in Ref. \cite{LA.11} that superheavy nuclei are very soft
objects: they possess very rich spectra of low-lying collective
vibrations (phonons). Therefore, correlations due to the
quasiparticle-vibration coupling (QVC) are then the next important
mechanism having considerable influence on the shell structure.
%

Medium-mass and heavy nuclei represent Fermi-systems where
single-particle and vibrational degrees of freedom are strongly
coupled. Collective vibrations lead to shape oscillations of the
mean nuclear potential and, therefore, modify the single-particle
motion. To take this effect into account, already in Ref.
\cite{BM.75} a general concept for the phonon coupling part of the
single-nucleon self-energy has been proposed. This concept has had
diverse implementations over the years
\cite{LR.06,SSV.80,MBBD.85,AK.99,BBG.99,B.09,CSFB.10,LA.11},
however, these studies either are not self-consistent or do not
include pairing correlations of the superfluid type. In this Letter
a fully self-consistent model, implementing both superfluid and
vibrational correlations in the relativistic framework, is
formulated and applied to a description of single-quasiparticle
spectra of tin and Z=120 isotopes.

%
%
Single-particle degrees of freedom in nuclei are characterized by
the single-(quasi)particle energies and the spectroscopic amplitudes
which can be determined in one-nucleon transfer or knockout
reactions. In microscopic many-body models these quantities enter
the well known Lehmann expansion of the one-body Green's function of
the N-body system over the eigenstates of the N$\pm$1-body systems
\cite{Mig.67}:
\be G(\xi,\xi^{\prime};\varepsilon) = \sum\limits_n \frac{
(\Psi(\xi))_{0n}(\Psi^{\dagger}(\xi^{\prime}))_{n0} } {\varepsilon -
(E_n^{(N+1)} - E_0^{(N)}) + i\delta} + \sum\limits_m \frac{
(\Psi^{\dagger}(\xi^{\prime}))_{0m}(\Psi(\xi))_{m0} } {\varepsilon +
(E_m^{(N-1)} - E_0^{(N)}) - i\delta}, \label{le} \ee \be
(\Psi^{\dagger}(\xi))_{n0} = \langle
\Phi_n^{(N+1)}|\Psi^{\dagger}(\xi)|\Phi_0^{(N)}\rangle, \ \ \ \ \ \
(\Psi(\xi))_{m0} = \langle
\Phi_m^{(N-1)}|\Psi(\xi)|\Phi_0^{(N)}\rangle, \ee
where $\delta\rightarrow+0$, $\Phi^{(N)}_0, \Phi^{(N)}_n$ are the
many-body wave functions of the ground and the excited state $n$ of
the N-body system, $E_0^{(N)}$, $E_n^{(N)}$ are its ground state and
excited state energies, and the variable $\xi$ includes the full set
of the single-particle variables in an arbitrary representation. The
numerators of Eq. (\ref{le})
give the spectroscopic amplitudes of the states $n$. To calculate
these spectroscopic amplitudes and the corresponding energies, the
Dyson equation is solved with a one-body Hamiltonian that consists
of a Relativistic Hartree-Bogoliubov (RHB) part $\mathcal{H}_{RHB}$
and an additional energy-dependent self-energy
$\Sigma^{(e)}(\varepsilon)$ \cite{LRT.08}:
\begin{equation}
\bigl(\varepsilon-\mathcal{H}_{RHB}-\Sigma^{(e)}(\varepsilon)\bigr)G(\varepsilon)=1.
\label{dyson}
\end{equation}
In the present work the space of the Dirac spinors diagonalizing the
RHB Hamiltonian is taken as a working basis. In this case $\xi =
\{k,\eta\}$ , where $k$ is the full set of the single-particle
quantum numbers in the spherical relativistic mean field (RMF) and
$\eta = \pm 1$ denotes the upper and lower components in the
Bogoliubov's quasiparticle space. Thus, the entities in the Eq.
(\ref{dyson}) are supermatrices in this space \cite{LRT.08}:
\be
\sum\limits_{\eta=\pm1}\sum\limits_{k}\Bigl((\varepsilon-\eta_{1}E_{k_{1}%
})\delta_{\eta_{1}\eta}\delta_{k_{1}k}-\Sigma_{k_{1}k}^{(e)\eta_{1}\eta
}(\varepsilon)\Bigr)G_{kk_{2}}^{\eta\eta_{2}}(\varepsilon)=
\delta_{\eta
_{1}\eta_{2}}\delta_{k_{1}k_{2}}. \label{fgdb}%
\ee
Here $E_k$ are the eigenvalues of the RHB hamiltonian and
$\Sigma_{k_{1}k}^{(e)\eta_{1}\eta }(\varepsilon)$ is the nucleonic
self-energy of the quasiparticle-phonon coupling \cite{LRT.08}:
\bea
\Sigma_{k_{1}k_{2}}^{(e)\eta_{1}\eta_{2}}(\varepsilon)=\sum\limits_{\eta=\pm
1}\sum\limits_{k,\mu}\frac{{\gamma}_{\mu;k_{1}k}^{\eta;\eta_{1}\eta}\ {\gamma}%
_{\mu;k_{2}k}^{\eta;\eta_{2}\eta\ast}}{\varepsilon-\eta
(E_{k}+\Omega_{\mu}-i\delta)}.
 \label{sfephon}%
\eea
The index $\mu$ labels the set of vibrational modes taken into
account and $\Omega_{\mu}$ are their frequencies computed with the
relativistic quasiparticle random phase approximation (RQRPA). The
vertices ${\gamma}$ determine the coupling of the quasiparticles to
the vibrational modes \cite{LRT.08}. In spherical nuclei, the
self-energy (\ref{sfephon}) has very small off-diagonal matrix
elements, thus the Green's function $G(\ve)$ is supposed to be
diagonal. Solutions of the Eq. (\ref{fgdb}) provide the
quasiparticle energies and the strength distributions (spectroscopic
factors) $S_n(\xi) = |(\Psi(\xi))_{n0}|^2$.

%
%
%
%
%
%

\begin{figure}[ptb]
\begin{center}
\includegraphics[scale=0.6]{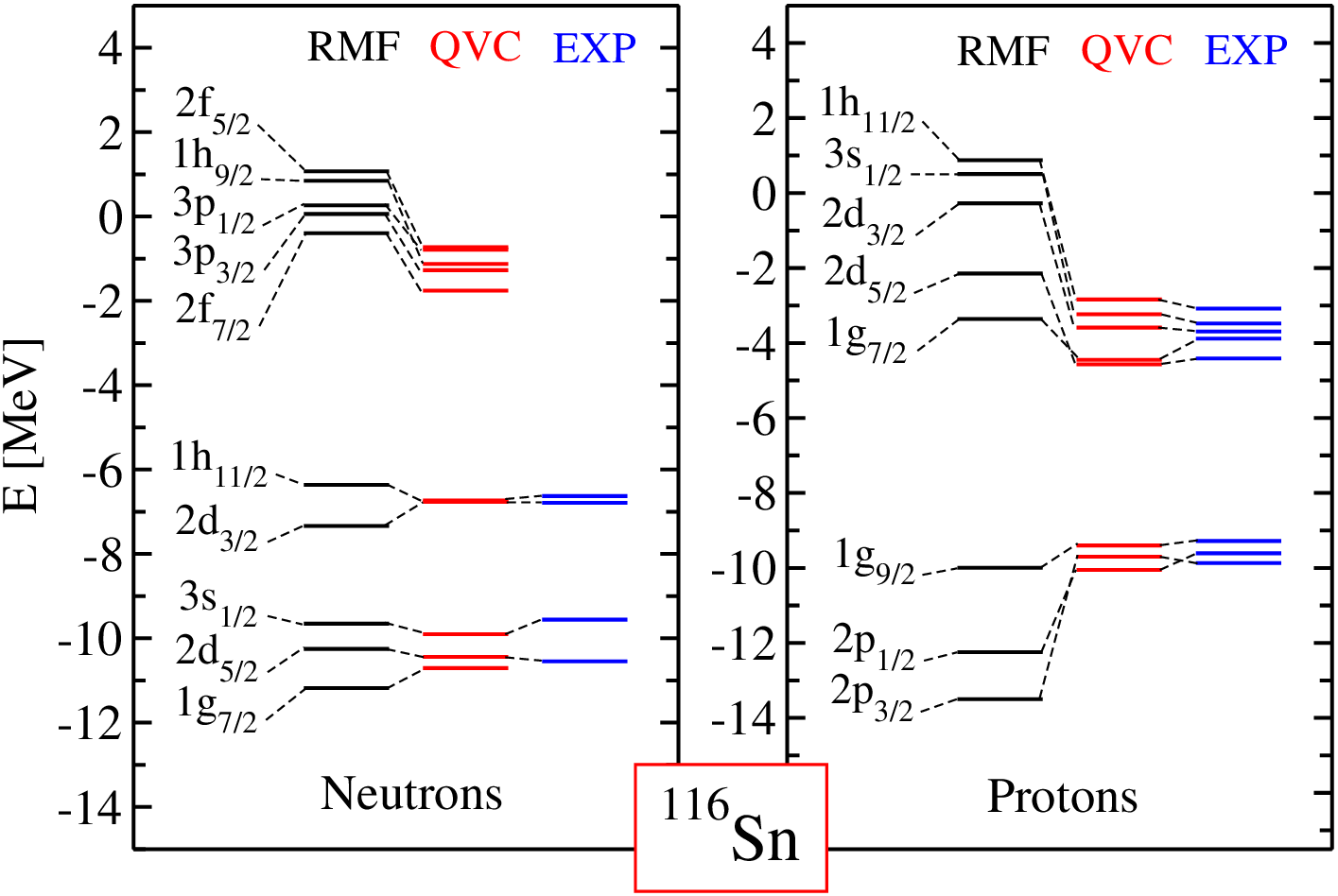}
\end{center}
\vspace{-5mm} \caption{(Color online) Single-quasiparticle spectrum
of $^{116}$Sn: RMF (left column), QVC (center) and experimental data
(right). In the 'QVC' and 'EXP' cases only the dominant levels are shown.}%
\label{116sn}%
\end{figure}

\begin{figure}[ptb]
\begin{center}
\includegraphics[scale=0.6]{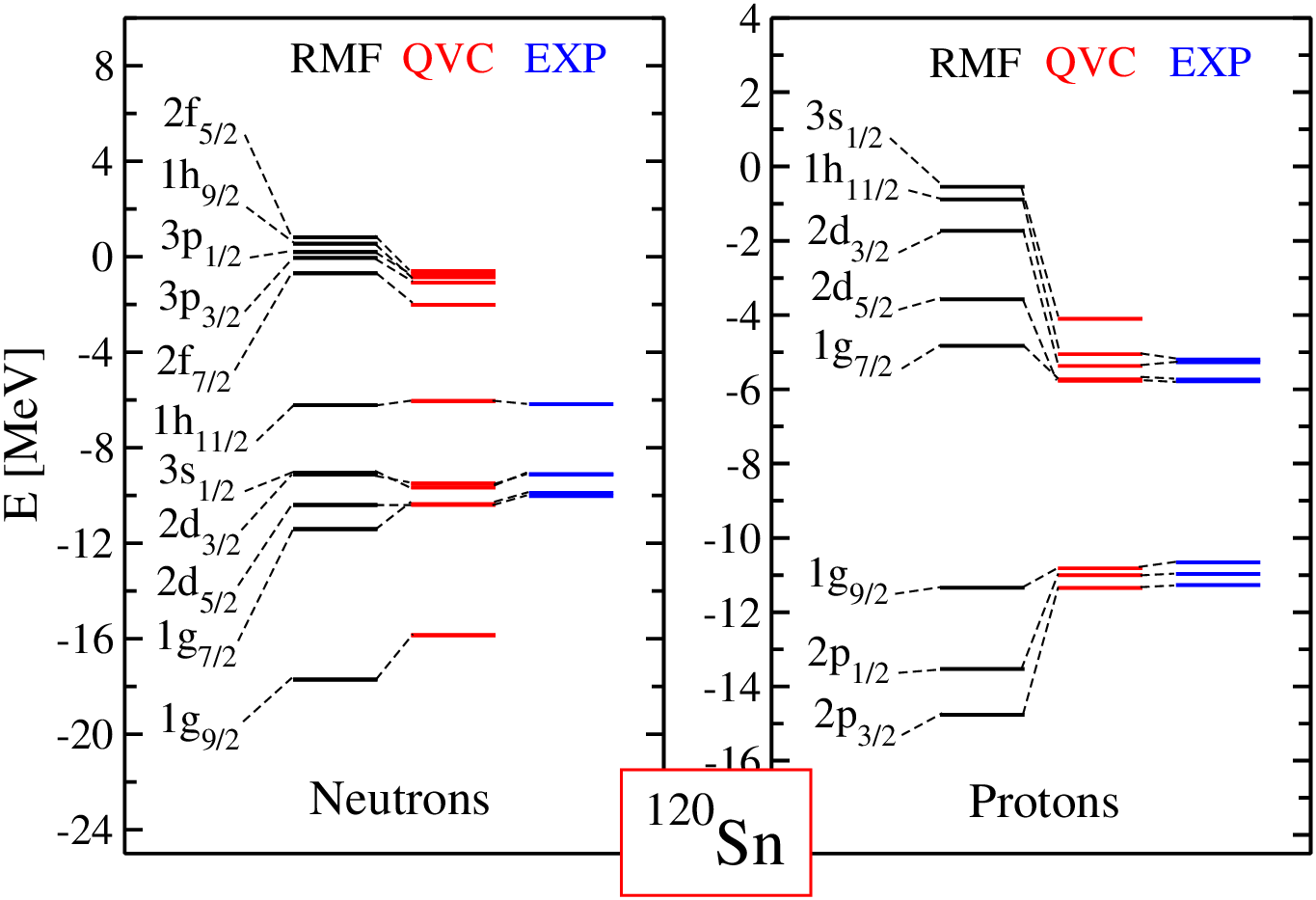}
\end{center}
\vspace{-5mm}
\caption{(Color online) Same as in Fig. \ref{116sn}, but for $^{120}$Sn.}%
\vspace{-5mm}
\label{120sn}%
\end{figure}

Before applying the model to unknown nuclei, benchmarking
calculations have been done for nuclei which had been investigated
experimentally. Tin isotopes represent a very good reference case as
their single-quasiparticle energies and spectroscopic factors in the
vicinity of the Fermi energy (FE) are known \cite{ENSDF}. The
neutron and proton single-quasiparticle levels in $^{116}$Sn and
$^{120}$Sn are shown in Figs. \ref{116sn} and \ref{120sn}. In each
panel, the left columns display the mean-field energies of the
Bogoliubov quasiparticles, the columns in the middle represent the
dominant levels (levels with the largest spectroscopic strength
$S_n(\xi)$) obtained within the QVC model and on the right the
experimentally observed dominant levels are shown.

In the applications to the doubly-magic nuclei \cite{LR.06,LA.11},
as a rule, the QVC correlations push the dominant levels towards the
FE. However, for the states, which are very close to the FE, the QVC
shift effect on the dominant levels is rather weak. In open-shell
neutron subsystems of $^{116,120}$Sn, where the FE is in the middle
of the shell, there are several states of this kind. Because of the
relatively high level density inside the valence shell, the leading
terms in the sum of Eq. (\ref{sfephon}) may compensate each other.
In contrast, in the closed-shell proton subsystems the level density
at the FE is smaller, thus, there are weaker compensations and the
shifts are considerably larger. For all dominant levels in
$^{116,120}$Sn nuclei one can find a very good agreement of the
presented QVC results with the data. The obtained spectroscopic
factors reproduce the available data also very well: their detailed
analysis will be presented elsewhere. The success of the response
theory built on the nucleonic self-energy of Eq. (\ref{sfephon})
\cite{LRT.08} can be now traced back to the results of the present
work showing that for a proper description of nuclear shell
structure and dynamics both types of correlations -- pairing and
quasiparticle-vibration coupling -- should be taken into account
self-consistently on the equal footing.

%
%
The $^{292,296,300,304}$120 isotopes have spherical minima of the
potential energy surfaces in both
Skyrme and CEDF calculations \cite{BBMR.04}, therefore, it is
justified to keep spherical symmetry for their mean-field
potentials. The newest non-linear meson-exchange interaction NL3*
\cite{NL3*} used for the CEDF in this work is the slightly improved
NL3 one \cite{NL3} known to give a very good agreement with data for
various low-energy phenomena not only in medium-mass nuclei, but
also in A$\simeq$250 mass region \cite{AAR.10} and, therefore, it is
justified to use this parameter set for the superheavy systems as
well as for the tin isotopes. The Bardeen-Cooper-Schrieffer model
for pairing correlations as well as the HFB model lead to a
collapsing solution in the Z=120 proton subsystem so that pairing
correlations can be restored only by approximate particle number
projection methods \cite{Af.06}. In the neutron subsystems, however,
no pairing collapse is found for Z = 120 nuclei, thus, in the
present work pairing has been included for the neutrons. In the
superheavy mass region the shell gaps are considerably smaller than
those between the previous shells. They amount about 2-3 Mev and,
therefore, are compatible with non-vanishing neutron pairing which,
in turn, slightly increases the gaps. The phonon spectra calculated
with the RQRPA in the chain of Z=120 isotopes show that these nuclei
are very soft: many rather collective phonons with $J^{\pi} = 2^{+},
3^{-}, 4^{+}, 5^{-}, 6^{+}$ are found below 15 MeV and included into
the self-energy (\ref{sfephon}).

\begin{figure}[ptb]
\begin{center}
\includegraphics[scale=0.65]{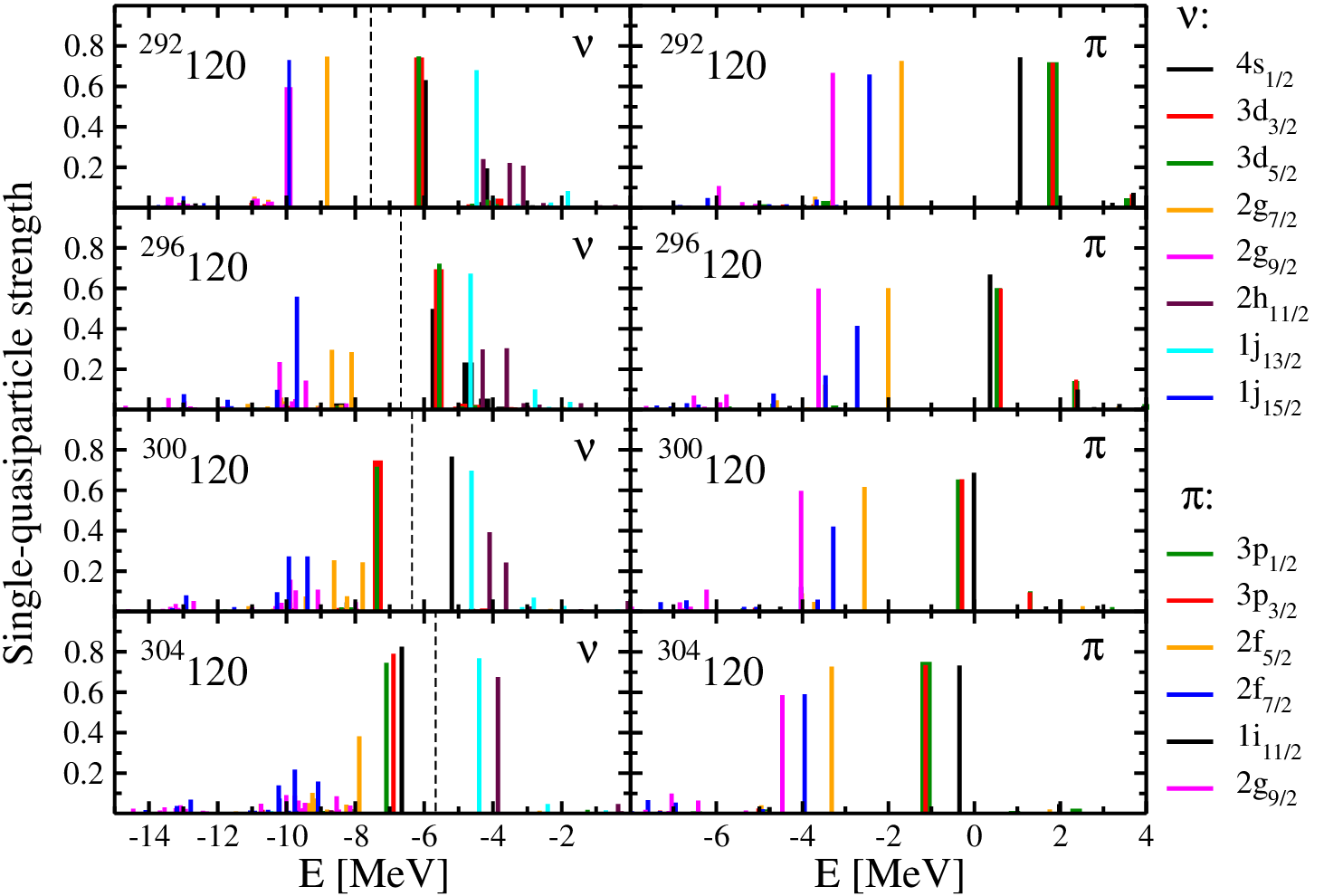}
\end{center}
\vspace{-5mm}
\caption{(Color online) Single-quasiparticle strength distribution
for the orbits around the Fermi surfaces in the neutron (left
panels) and proton (right panels) subsystems of the Z=120 isotopes
calculated in the relativistic quasiparticle-vibration coupling model.
The dashed lines indicate the chemical potentials.}%
\vspace{-5mm}
\label{z120}%
\end{figure}
The selected results on the single-quasiparticle strength
distributions in the neutron and the proton subsystems of the Z =
120 isotopic chain are displayed in Fig. \ref{z120}. The
distributions for the orbits closest to the neutron and the proton
FE's are given and denoted by different colors. Thus, one can see
the evolution of these distributions with an increase of the neutron
number from N = 172 to N = 184. As in the neutron subsystems both
pairing and QVC mechanisms are included, their very delicate
interplay is found: pairing correlations tend to increase the shell
gap while the QVC tends to decrease it and at the same time causes
the fragmentation of the states in the middle of the shell. As a
result, in the presence of both mechanisms the gap in the neutron
subsystem remains almost steady while the newly occupied levels jump
down over the gap when the neutrons are added. The shell gap in the
proton subsystems of the considered nuclei diminishes only slightly
when the neutron number increases, so that the proton number Z = 120
remains a rather stable shell closure while the detailed structure
of the proton levels shows some rearrangements induced by the
neutron addition.


\begin{table}[ptb]
\caption{Alpha decay energies [MeV] for even-even Z=120 isotopes
with N = 176-184 calculated in the covariant density functional
theory without (RMF) and with (RMF+VC) vibrational corrections.}
\tabcolsep=1.0em \renewcommand{\arraystretch}{1.0}%
\begin{tabular}
[c]{cccccc} \hline\hline \
N & 176 & 178 & 180 & 182 & 184 \\
\hline RMF & 11.81 & 11.55 & 11.28 & 11.23 & 11.60\\
RMF+VC & 11.73 & 11.68 & 11.65 & 11.41 & 11.91\\
\hline\hline \label{tab}
\end{tabular}
\vspace{-8mm}
\end{table}

The zero-point fluctuations associated with the nuclear vibrational
motion affect the nuclear binding energies \cite{RS.80,BBB.06}. This
effect is especially important for the superheavy alpha-emitters as
their lifetimes are related directly to the differences of the
binding energies of the mother and the daughter nuclei. In order to
obtain the correct nuclear binding energy, the vibrational
correlations should be, in principle, incorporated into the fitting
procedure for the underlying energy density functional
\cite{BBB.06}. However, for the differential quantities like alpha
decay energies the vibrational corrections (VC) can be performed
after the minimization procedure. For all the considered Z = 120
isotopes the RQRPA vibrational corrections to the total energies
amount 4-5 MeV being comparable to the shell correction energies.
Alpha decay energies for the even-even Z=120 isotopes calculated
within the RMF and the RMF+VC models are displayed in Table
\ref{tab}. One can see that the vibrational corrections to the alpha
decay energies reach several hundred keV and can be both positive
and negative introducing either stabilizing or destabilizing effect.
As a result, the lifetime predictions change correspondingly: in the
case of the largest VC in $^{300}$120, the decrease of the lifetime
of this nucleus is up to an order of magnitude, depending on the
evaluation method \cite{DK.09}.

%
%
In conclusion, the relativistic quasiparticle-vibration coupling
model is formulated. The results obtained for the experimentally
known nuclei illustrate that the self-consistent implementation of
many-body correlations beyond the CEDF represents a successful
strategy towards a universal and precise approach for the low-energy
nuclear dynamics. The model has allowed looking deep inside the
shell structure of Z = 120 isotopes representing hypothetically an
island of stability for superheavy nuclei. It has been found that
the proton number Z = 120 remains a rather stable spherical shell
closure when the neutron number changes from N = 172 to N = 184. In
the neutron subsystem, due to an interplay of pairing and
quasiparticle-vibration coupling, the smooth evolution of the shell
structure is observed, so that at all the neutron numbers N =
172,176,180,184 comparable spherical shell gaps are found. The
analysis of the alpha decay energies has included, for the first
time, the vibrational corrections and shown that these corrections
can amount several hundred keV in both directions smearing the
irregularities due to the shell effects.

%
%
Valuable discussions with A. Afanasjev, H. Feldmeier, P. Ring, V.
Tselyaev are gratefully acknowledged. This work was supported by the
LOEWE program of the State of Hesse (Helmholtz International Center
for FAIR) and the Alliance Program of the Helmholtz Association
(HA216/EMMI).


\end{document}